\documentclass{iau}
\usepackage{graphicx} 
\usepackage{amsmath,amssymb,latexsym}
\title[IAUS291.~~The All-Sky High Time Resolution Universe Survey]
{Conducting The Deepest All-Sky Pulsar Survey Ever: The All-Sky High Time Resolution Universe Survey}

\author[Ng \& the HTRU Collaboration] 
{Cherry Ng$^1$
 \and the HTRU Collaboration}

\affiliation{$^1$Max-Planck-Institut f\"{u}r Radioastronomie, \\ Auf dem H\"{u}gel 69, 53121 Bonn, Germany \\ email: {\tt cherryng@mpifr-bonn.mpg.de} }

\pubyear{2012}
\volume{291} 
\jname{\mbox{Neutron Stars and Pulsars: Challenges and Opportunities after 80 years}}
\editors{J. van Leeuwen, ed.} 
\begin{document}

\maketitle

\begin{abstract}
The extreme conditions found in and around pulsars make them fantastic natural laboratories, providing insights to a rich variety of fundamental physics and astronomy. To discover more pulsars we have begun the High Time Resolution Universe (HTRU) survey: a blind survey of the northern sky with the 100-m Effelsberg radio telescope in Germany and a twin survey of the southern sky with the 64-m Parkes radio telescope in Australia. The HTRU is an international collaboration with expertise shared among the MPIfR in Germany, ATNF/CASS and Swinburne University of Technology in Australia, University of Manchester in the UK and INAF in Italy. The HTRU survey uses multi-beam receivers and backends constructed with recent advancements in technology, providing unprecedentedly high time and frequency resolution, allowing us to probe deeper into the Galaxy than ever before. While a general overview of HTRU has been given by Keith at this conference, here we focus on three further aspects of HTRU discoveries and highlights. These include the `Diamond-planet pulsar' binary PSR~J1719-1438 and a second similar system recently discovered. In addition, we provide specifications of the HTRU-North survey and an update of its status. In the last section we give an overview of the search for highly-accelerated binaries in the Galactic plane region. We discuss the computational challenges arising from the processing of the petabyte-sized HTRU survey data. We present an innovative segmented search technique which aims to increase our chances of discovering highly accelerated relativistic binary systems, potentially including pulsar-black-hole binaries.
\end{abstract}

\firstsection
             
\section{`Planet-pulsar' binaries}
In 2009 HTRU discovered PSR~J1719-1438, a 5.7-millisecond pulsar (MSP), with the Parkes 64-meter radio telescope. After follow-up timing it was shown that this pulsar is in a binary system with an orbital period of 2.2 hours. The mass of the companion has been found to be 0.00115\,M$_{\odot}$, which is comparable to that of Jupiter ($\sim$1.2\,M$_{\rm{J}}$). However with a minimum density of 23\,g\,cm$^{-3}$, this companion has physical properties which are clearly incompatible with those of Jupiter ($\rho$$_{\rm{J}}$$\textless$\,2\,g\,cm$^{-3}$). PSR~J1719-1438 does not show signs of solid-body eclipses, making it unlikely to be a black widow system (\cite[Fruchter, Stinebring \& Taylor 1988]{FST88}). 

Further optical observations with the Keck 10-m Telescope revealed that the companion is likely to be an ultralow-mass carbon white dwarf that has lost 99\% of its mass, possibly the remains of the degenerate core of the original white dwarf. Ultracompact low-mass x-ray binaries are potential progenitors of this system, in which the companion has narrowly escaped complete destruction. In fact under such conditions carbon would be crystallised, hence the nickname `Diamond planet' (\cite[Bailes et al. 2011]{B11}).

A second such `planet-pulsar' system has recently been discovered in the HTRU survey. This second system has parameters very similar to those of PSR~J1719-1438 (Table 1), with a companion mass of the order of magnitude of a planet. Further timing studies are currently underway to improve the phase-coherent timing solution (\cite[Thornton et al. in prep]{Thornton}). These HTRU results may be shedding light on a previously unknown population of pulsars, and perhaps more of such systems will be discovered in the near future.  
\begin{table}[htpb]
\caption{Comparison between the two `Planet-pulsar' binaries.}
\begin{center}
\begin{tabular}{lll}
\hline
    & J1719-1438 & `Planet-pulsar' n$^{\circ}$2 \\ \hline \hline \\ [-2.0ex]
   P$_{\rm{spin}}$ (ms) & 5.7 & 3.46 \\
   DM Distance (kpc) & 1.2(3) & 0.313 \\
   P$_{\rm{orb}}$ (days) & 0.09 & 0.32 \\
   a $\sin${\it i} (lt-s) & 0.0018 & 0.002 \\

   m$_{\rm{comp}}$ (M$_{\odot}$) & $\geqslant$ 0.00115~($\sim$1.2 M\,$_{\rm{Jovian}}$) & $\geqslant$ 0.00076~($\sim$0.8\,M$_{\rm{Jovian}}$) \\
\hline
\end{tabular}
\end{center}
\end{table}
\vspace{-0.6em}
\section{HTRU-North survey with the Effelsberg Telescope}
The ongoing HTRU-North survey is being conducted using the 100-m Effelsberg radio telescope operated by the Max-Planck-Institut f\"{u}r Radioastronomie. HTRU-North is the first large-scale pulsar survey performed with the Effelsberg telescope. Most importantly, the survey includes the region north of 38$^{\circ}$ Galactic latitude, which has remained unsurveyed for the last 20 years. As a twin to the HTRU-South survey, the integration lengths at Effelsberg have been scaled down to achieve the same sensitivity as that of the Parkes Telescope. See Table 2 for a summary of the specifications for both surveys.
\begin{table}[htpb]
\caption{Summary of the specifications of the HTRU-North and HTRU-South surveys.}
\begin{center}
\begin{tabular}{p{2.4cm}p{5.1cm}p{5.3cm}}
\hline
 & Northern survey & Southern survey \\ \hline \hline \\ [-2.0ex]
Start date   & Summer 2010 & Early 2008 \\ 
Telescope    & Effelsberg-100m & Parkes-64m \\ 
Sky coverage & $\delta$ > -20$^{\circ}$ & $\delta$ < +10$^{\circ}$ \\ 
T$_{\rm{obs}}$ (s)   & Low-lat: 1500   & Low-lat: 4300 \\ 
              & Mid-lat: 180    &  Mid-lat: 540  \\
              & High-lat: 90    & High-lat: 270  \\
Receiver     & 7-beam 1.4-GHz receiver & 13-beam 1.35-GHz receiver \\ 
Backend      & Pulsar Fast Fourier Transform Spectrometer (PFFTS) & Berkeley-Parkes-Swinburne Recorder (BPSR) \\ 
{\it B} (MHz)      & 300 & 340 \\ 
{\it N}$_{\rm{chans}}$    & 512 & 1024 \\ 
$\Delta$ $\nu$$_{\rm{chan}}$ (MHz) & 0.58 & 0.39 \\ 
T$_{\rm{samp}}$ ($\mu$s) & 54 & 64 \\ 
N$_{\rm{pointings}}$ & $\sim$180,000 & $\sim$43,000 \\ 
Total data (PB) & $\sim$5 & $\sim$1  \\ 
\hline 
\end{tabular}
\end{center}
\end{table}

Observations of unidentified Fermi point sources were taken initially as a system test for the HTRU-North survey. This has led to the discovery of Effelsberg's first ever MSP~J1745+1017 (\cite[Barr et al. submitted]{Barr}). The HTRU-North survey officially began in 2010. Thus far, $\sim10\%$ of the medium-latitude pointings have been processed. Already 12 pulsars have been discovered, highlights of which include a MSP~J1946+3414 in a highly eccentric orbit and a young pulsar PSR~J2004+3427 with a characteristic age $\le$19\,kyr (\cite[Barr et al. in prep]{Barr}). Extrapolating from these statistics we expect to find $\sim$100 new pulsars from the medium-latitude region alone. Population synthesis studies predict that of the order of 500 new pulsars are to be discovered from the complete HTRU-North survey, which includes $\sim$40 new MSPs.

\section{Searching for highly-accelerated binaries in the Galactic plane}
Pulsars in tight binary orbits with other compact objects such as neutron stars and, potentially, black holes are of great interest as their strong gravitational fields provide the best tests for General Relativity (GR) and other theories of gravity. The best example of such a binary system so far is the Double Pulsar (\cite[Burgay et al. 2003]{B03}, \cite[Lyne et al. 2004]{L04}). The Double Pulsar has been used to obtain four independent tests of GR predictions and has shown that GR passes these yet most stringent tests with a measurement uncertainty of only 0.05\% (\cite[Kramer et al. 2006]{K06}). In fact, deliverable science scales directly with the compactness of the binary system to be discovered and our aim is the discovery and study of ultra-compact relativistic binary systems. 

The deep Galactic-Plane survey is where the most relativistic binaries are expected to be found (\cite[Belczynski et al. 2002]{B02}). In this region of sky within Galactic latitude $\pm$3.5$^{\circ}$, we employ the longest integrations to maximise our sensitivity (see Table 2). However, the sheer volume of data poses great challenges in data manipulation and analysis. From the southern sky alone we expect to have about 250\,TB of data. When it comes to searching for binary pulsars, the analysis procedure becomes very complicated, since the orbital parameters of the binary system must be taken into account, resulting in a large 7-dimensional parameter space. We have recently implemented an innovative pipeline targeting the search of binary systems. Our strategy is to approximate any unknown orbital motion as a simple line-of-sight acceleration. We split these long data sets into sections and search for binary systems using an acceleration range appropriate for the length of the section. We push the maximum achievable acceleration value to over 1200\,m/s$^2$. In other words, within this acceleration range we can find a Double-Pulsar-like system deeper in the Galaxy with orbital periods shorter than 1 hour and would also be able to explore the parameter space occupied by expected pulsar-black-hole systems.

In order to probe the full parameter space for binary pulsars, we compute at least 1.46 million Fourier transform operations per data set. Such ‘acceleration searches’ are obviously computationally very expensive and for example would have taken a single-CPU computer over 1400 years to completely process just the Southern Galactic Plane data. We are carrying out processing across multiple international high-performance computer clusters. As a system test we have first analysed the HTRU data set containing the Double Pulsar. We report a three-fold increase in sensitivity from the analysis with acceleration search compared to the analysis without (Figure 1), illustrating the importance of this binary search pipeline.
\begin{figure}[hbp]
\begin{center}
\includegraphics[width=3.9in]{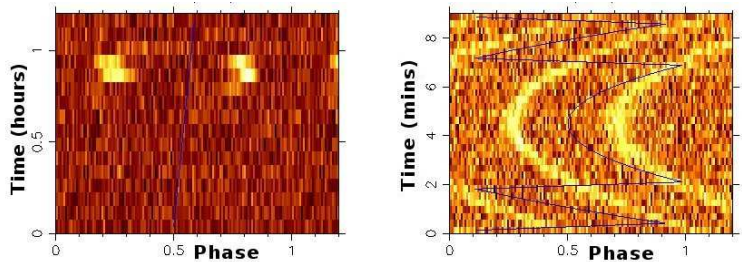}
\caption{Observation time versus pulse phase plots for the Double Pulsar data set. (Left) Analysis without acceleration search on the full 72-min data results in a signal-to-noise ratio (SNR) of 11.73, only marginally above the detection limit. (Right) Analysis with acceleration search on a 9-min section results in an improved SNR of 31.79. The orbital motion of the Double Pulsar is clearly seen, as traced out by the blue line indicating the best fit model.}
\end{center}
\end{figure}

\section{Conclusion}
The HTRU-North survey with the Effelsberg Telescope began in 2010. Currently 20\% of the medium-latitude and 5\% of the high-latitude region have been observed, and $\sim$10\% of the medium-latitude data have been processed. This has led to the discovery of 13 new pulsars, including one MSP and one Fermi MSP (Figure 2). For the HTRU-South survey, observations of the medium-latitude region have been completed. 40\% of the high-latitude and more than 80\% of the low-latitude region have been observed. Over 120 new pulsars have been found in the HTRU-South survey, including 27 MSPs (Figure 2). Highlight discoveries are for example the magnetar PSR~J1622-4950 (\cite[Levin et al. 2010]{L10}), `planet-pulsar' binaries as mentioned in \S1 and transient objects (\cite[Burke-Spolaor et al. 2011]{BS11}). The HTRU low-latitude data will provide the deepest large-scale search thus far in the Galactic plane region. Up to now we have processed, without acceleration searches, 15\% of the Galactic Plane data from the southern sky. Already we have discovered 23 new pulsars and redetected about 200 known pulsars. An innovative `acceleration search' pipeline has been implemented recently, targeting the search for binary systems in the low-latitude data. This processing will surely return even more exciting new discoveries.  
\begin{figure}[hbpt]
\begin{center}
 \includegraphics[width=4.0in]{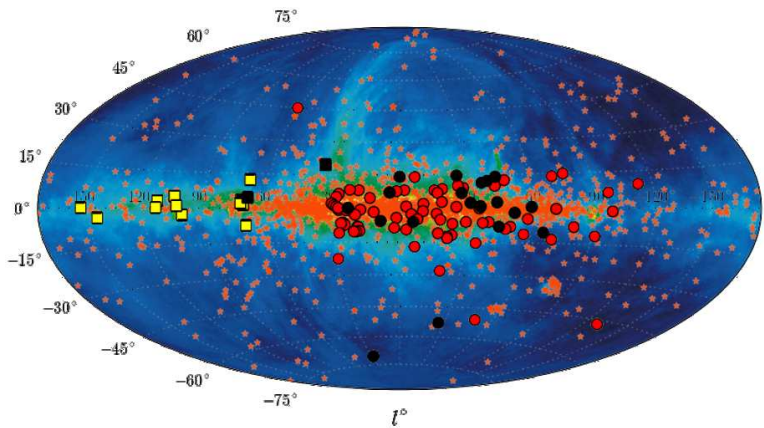}
 \caption{Up-to-date HTRU discoveries overplotted on the \cite[Haslam et al. (1982)]{H82} sky map. Circles indicate new HTRU-South pulsars (red: normal pulsars, black: MSPs). Squares indicate new HTRU-North pulsars (yellow: normal pulsars, black: MSPs). Orange stars indicate previously known pulsars.}
\end{center}
\end{figure}

\end{document}